\documentclass{article}
\usepackage{ijcai23}

\usepackage{times}
\usepackage{soul}
\usepackage{url}
\usepackage[hidelinks]{hyperref}
\usepackage[utf8]{inputenc}
\usepackage[small]{caption}
\usepackage{graphicx}
\usepackage{amsmath}
\usepackage{amsthm}
\usepackage{amssymb}
\usepackage{booktabs}
\usepackage{algorithm}
\usepackage{algorithmic}
\usepackage[switch]{lineno}
\usepackage{diagbox}
\usepackage{multirow}
\usepackage{subfigure}
\usepackage{float} 

\usepackage{xcolor}

\newcommand{\ignore}[1]{{}}

\newtheorem{definition}{Definition}

\title{Privacy Protectability: An Information-theoretical Approach}

\author{
Siping Shi$^1$
\and
Bihai Zhang$^1$\and
Dan Wang$^{1}$
\affiliations
$^1$The Hong Kong Polytechnic University
\emails
\{cssshi, csbzhang, csdwang\}@comp.polyu.edu.hk
}

\begin{document}

\maketitle

\begin{abstract}
Recently, inference privacy has attracted increasing attention. The inference privacy concern arises most notably in the widely deployed edge-cloud video analytics systems, where the cloud needs the videos captured from the edge. The video data can contain sensitive information and subject to attack when they are transmitted to the cloud for inference. Many privacy protection schemes have been proposed. Yet, the performance of a scheme needs to be determined by experiments or inferred by analyzing the specific case. In this paper, we propose a new metric, \textit{privacy protectability}, to characterize to what degree a video stream can be protected given a certain video analytics task. Such a metric has strong operational meaning. For example, low protectability means that it may be necessary to set up an overall secure environment. We can also evaluate a privacy protection scheme, e.g., assume it obfuscates the video data, what level of protection this scheme has achieved after obfuscation. Our definition of privacy protectability is rooted in information theory and we develop efficient algorithms to estimate the metric. We use experiments on real data to validate that our metric is consistent with empirical measurements on how well a video stream can be protected for a video analytics task.
\end{abstract}

\section{Introduction}

Inference privacy is an important research area of privacy-preserving machine learning (PPML) and has attracted increasing attention recently~\cite{xu2021privacy}. In practice, inference privacy arises from the widely deployed video analytics systems, e.g., surveillance, autonomous driving, Metaverse, etc. Such a system is usually based on an edge-cloud architecture. The edge systems capture video stream data. The cloud has pre-trained models and, due to the resource constraints in the edge or the video analytics task requiring edge collaboration, the cloud needs the edge systems to transmit their videos or intermediate analytics results to complete the inference. The videos captured in the edge can contain sensitive information, e.g., humans, license plates, etc., and such information can be subject to attack in transmission.

To protect privacy, a secure environment can be set up using encryption-based schemes~\cite{pmlr-v48-gilad-bachrach16,pmlr-v139-li21e} or trusted environment-based schemes~\cite{narra2019privacy,Bian21}. Yet these schemes bring about significant overheads. Recently, research efforts explore privacy protection schemes based on perturbation, i.e., those obfuscate the video data through adding noise \cite{Mireshghallah2020}, object removal \cite{Bihai2022}, image transformation \cite{Gao2021}, differential privacy \cite{NhatHai2016}, etc. These schemes have much fewer resource footprints and are practical for edge systems. We see adoption and deployment \cite{google}. 

\ignore{
\begin{figure}[t!]
	\centering 
	\includegraphics[width=0.95\linewidth]{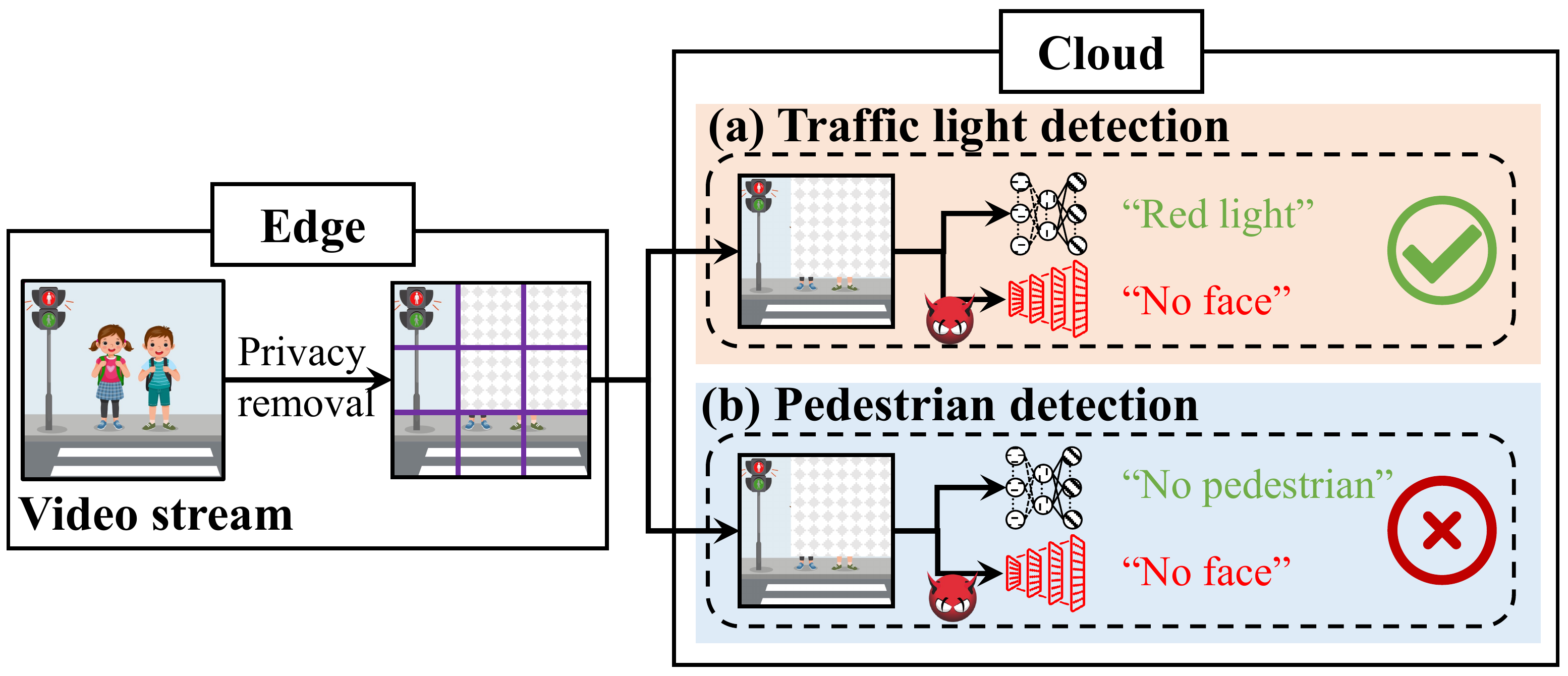}
	\caption{An illustration of privacy protectability.}
\label{protectability-moti}
\end{figure}
}


However, the choice of a protection scheme and its configurations are usually determined by experiments or inferred by analyzing the specific case.
There are reports that a protection scheme \cite{Singh2021} may not work. The imperative question is that \textit{when a privacy protection scheme works and to what extent it works}. In this paper, we estimate this \textit{analytically} by proposing a new metric, \textit{privacy protectability}, to characterize to what degree a video stream can be protected given a certain video analytics task. We observe that, intuitively, a video stream with sensitive information, e.g., humans, 
may be easier to get protected for a video analytics task of traffic light detection, while it may be difficult to get protected for a video analytics task of pedestrian detection. 


Intrinsically, there are sensitive features in a video stream and these features may also contribute to the accuracy of the video analytics task. Their overlap determines the protectability. We develop a privacy protectability metric, $\mathcal{P}$-score, through statistical and information theory. The key is to quantify how much each feature contributes to the accuracy of the video analytics task and how much each feature contributes to the accuracy of a privacy attack, e.g., the typical attribute inference attack \cite{Shagufta2022,Gong2018}. We resort to predictive power \cite{Covert2020,pmlr-v139-catav21a} to quantify such contribution. We call a feature a \textit{protectable feature} if its contribution to the sensitive attribute is less than a threshold. Intuitively, a protectable feature will not release enough sensitive information to a privacy attack. We then define \textit{privacy protectability} as the ratio of protectable features to the whole feature sets. We develop an algorithm to efficiently compute privacy protectability through approximation and a sample-based method.

Our metric has strong operational meanings. For example, low protectability means that a perturbation-based privacy protection scheme will not work, i.e., it is not possible to both protect the privacy of the video stream and achieve high accuracy of the video analytics task after perturbation. One can either resort to the schemes that set up a secure environment or use another video stream for the analytics task. High protectability means that the video stream is \textit{protectable}, i.e., the sensitive features can be separated without affecting the analytics accuracy significantly. To ensure that the video stream is \textit{protected} after applying a privacy protection scheme, we further develop a level-of-protection metric, a $\mathcal{LP}$-score and its estimation algorithm. The $\mathcal{LP}$-score can assist in evaluating the appropriate choices of the perturbation-based protection schemes and their configurations.

We evaluate privacy protectability with real-world data. We show that our new metrics are consistent with empirical measurements on how well a video stream can be protected and our metrics can be computed efficiently. Our metrics have strong operational meaning. Currently, judging whether a video stream for a video analytics task can be protected is a case-by-case effort. It requires applying privacy protection schemes and conducting video analytics, which is resource-consuming. Our metric provides a definitive judgment and our sample-based algorithm can be efficiently computed. We present a case study to show how our metrics can be used by practitioners. Specifically, we study two video analytics tasks, a traffic light detection task and a crowd density detection task. We study multiple video streams and which of them are protectable when serving video analytics tasks. For those that are not protectable, we apply three state-of-the-art privacy protection schemes and show that indeed they do not perform well. For those protectable, we apply protection schemes and examine the level-of-protection under different configurations. We show that our metrics can be efficiently computed to facilitate these evaluations. As a comparison, a case-by-case evaluation of privacy protection schemes requires more than two orders of resources.

The contributions of this paper are: 

\begin{itemize}

\item We propose a new metric privacy protectability to evaluate to what degree a video stream can be protected given a certain video analytics task. As we see an increasing number of privacy protection schemes developed, the privacy protectability metric can help to judge when a privacy protection scheme may work and to what extent.

\item The design of our new metric has salient features: (1) it is theoretically driven and has a strong operational meaning rooted in statistics and information theory; (2) it can be computed directly and efficiently from the input data, with fewer samples than those needed for empirical learning; and (3) it can be shown to be strongly consistent with empirical measurements.

\item We evaluate our metric with real-world data and we demonstrate how our metric can be used through a case study in edge-cloud video analytics systems.

\end{itemize}

\ignore{
\section{Motivation Example}

In a marathon, various analytics tasks are performed for the security and safety of all participants based on the videos captured by deployed cameras. For example, the crowd analysis task is executed to avoid stampede accidents, by automatically counting people from the captured videos in real time. The facial emotion recognition task is also performed to monitor the health status of the participants, and advise the participant to rest if the pain emotion is detected on their face. Particularly, for the privacy of the participants, the sensitive ethnic group information should not be exposed during analyzing. To achieve this, several privacy schemes are required to perform for the analytics task. Since the difficulty of protecting privacy for each analytics task is various, different privacy schemes can be applied to different tasks. Privacy protectability is proposed to evaluate the difficulty of privacy protection. Specifically, analytics task with higher privacy protectability (i.e., the crowd analysis task) indicates less difficulty of privacy protection, and a simple privacy scheme (i.e., sensitive feature removal method) can be applied to achieve the privacy preserving goal. Otherwise, analytics task with lower privacy protectability (i.e., the facial emotion recognition task) indicates more difficulty of privacy protection, and a complex privacy scheme (i.e., sensitive feature encryption) is required. 

}

\section{Preliminaries}
We first give the notations and the threat model in video analytics. We then introduce predictive power~\cite{Covert2020,pmlr-v139-catav21a}, a measure of information. Predictive power is developed in the machine learning model explanation area in the efforts towards feature attribution, i.e., quantifying the importance of a feature in machine learning model prediction. Predictive power is a recent progress to quantify the contribution of a feature subset for predicting the target attribute with the machine learning model.

\textbf{Notations.} We consider a video analytics service. Let $X$ be a video stream captured by the service consumer and analyzed by the service provider. Let $Y_a$ be a target analytics attribute of $X$, e.g., traffic light, and $Y_{pri}$ be the private information of $X$, e.g., human face. Let $F$ be the pre-trained deep neural network (DNN) model for the video analytics task from the service provider. Let $Z= \{z_1, z_2, \ldots, z_n\}$ be a set of extracted features of $X$ for model $F$, e.g.,  red pixels, edges of face. Here $n$ is the number of features, and $z_i$ is the $i$-th individual feature.

\textbf{Threat model.} In this paper, we assume the service provider and the transmission environment is untrusted, and the privacy attacker could be any party who has an interest in the sensitive information of the transmitted data from the service consumer. The attacker is assumed to have unbounded computation capability and can hijack the data transmitted by the service consumer to analyze sensitive information.

\textbf{Predictive power.} Let $S$ be a subset of the full feature set $Z$. We define $F_S$ as the prediction model restricted to using only feature set $S$ to predict $Y_a$. There are two special cases where $S=\emptyset$ and $S=Z$, which respectively correspond to the mean prediction model $F_{\emptyset}$ and the full prediction model $F$. Let $\ell(\cdot)$ be the loss function, and $\mathbb{E}[\ell(F_S(X), Y_a)]$ be the expected loss taken over the data distribution $p(X, Y_a)$ with model $F_S$. Formally, the predictive power of feature set $S$ is defined as the expected loss reduction between the mean prediction model $F_{\emptyset}$ and the feature set $S$ restricted model $F_S$, which is given by: 
\begin{equation}
    \begin{aligned}
        \nu_F(S,Y_a)=\mathbb{E}[\ell(F_{\emptyset}(X), Y_a)]-\mathbb{E}[\ell(F_S(X), Y_a)].
    \end{aligned}
\end{equation}
\section{Privacy Protectability}

\subsection{Overview}
\begin{figure}[t!]
	\centering 
	\includegraphics[width=0.98\linewidth]{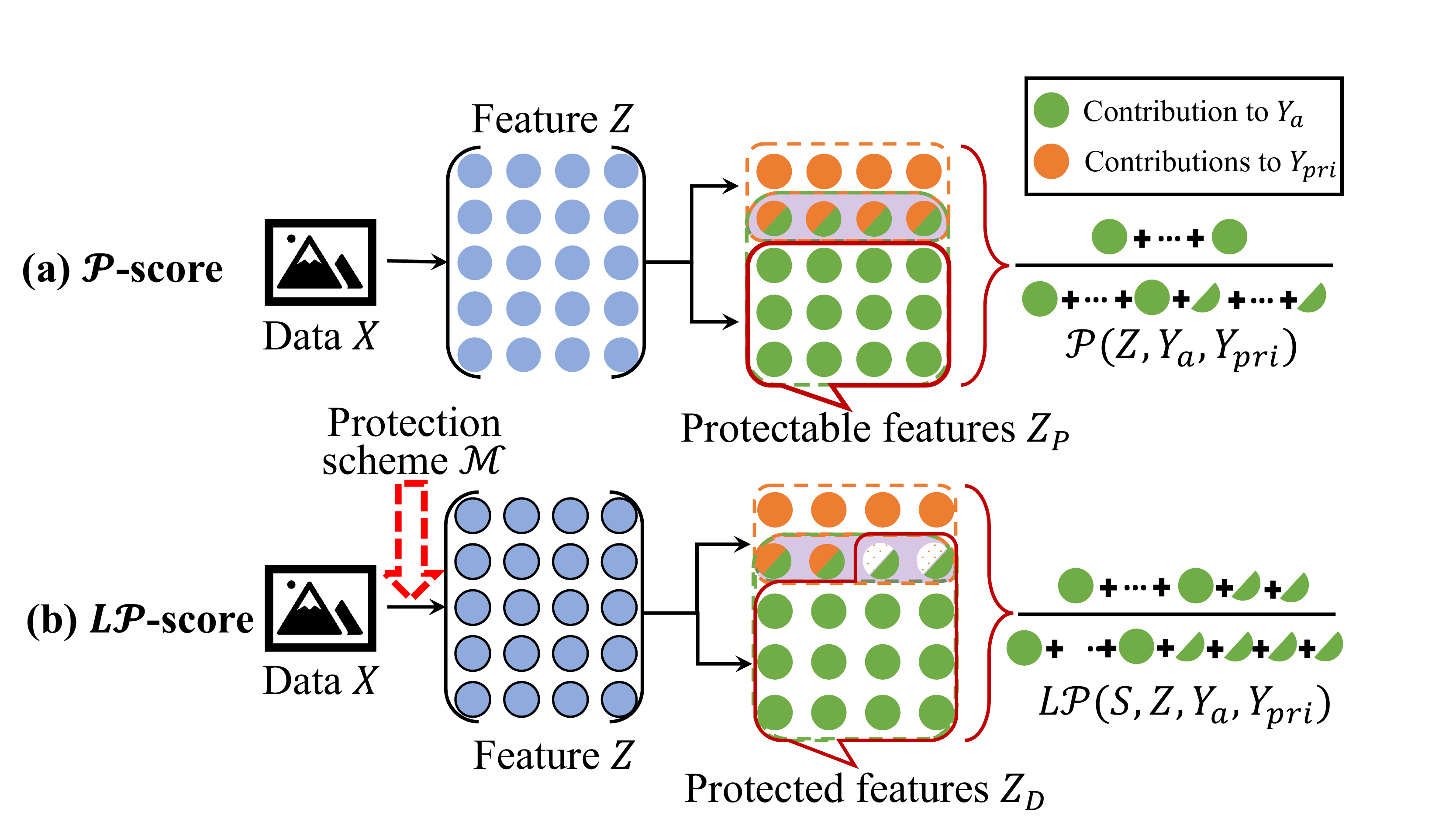}
	\caption{An overview of privacy protectability definition.}
\label{protectability-demo}
\end{figure}

We use Fig. \ref{protectability-demo} to illustrate an overview of privacy protectability.
Intrinsically, a video analytics model $F$ makes decisions based on the input feature set $Z$ extracted from the video stream $X$. In other words, $Z$ contributes to predicting the attribute $Y_a$. We will characterize a feature contribution score of each feature (Section~\ref{sec_protectable}) through its mean predictive power to $Y_a$; we will use a Shapely Value function to define the expected mean predictive power over the predictive power of all feature combinations. In $Z$, some features may be difficult to protect as they contain sensitive information. In other words, they contribute to predicting the privacy attribute $Y_{pri}$. Accordingly, we will define (Section~\ref{sec_protectable}) protectable features $Z_P$. Intuitively, the more protectable features, the higher probability that the video stream can be protected. We will define privacy protectability (Section~\ref{sec_protectability}) to be the ratio of protectable features against the full feature set. We define protectability through a ratio because if the ratio of protectable features is high, it is possible to achieve a high analytics accuracy even if we perturb unprotectable features. Fig. \ref{protectability-demo}(a) shows the $\mathcal{P}$-score, $\mathcal{P}(Z, Y_a, Y_{pri})$, as a function of a video stream extracted features $Z$, a video analytics task represented by $Y_a$, privacy attribute $Y_{pri}$. The computation of privacy protectability is exponential, due to the computation of the predictive power and the Shapely Value function. Thus, we will develop an efficient algorithm (Section~\ref{sec_protectability}) based on an approximation of the predictive power and a sampling method for the Shapely Value function. 

We comment that privacy protectability is a measure to quantify whether a video stream $X$ for a video analytics model $F$ (representing the video analytics task) can be protected, i.e., high protectability means that the video can be perturbed in a way with high analytics accuracy and low privacy leakage, and vice versa for low protectability. In practice, given high protectability, we still need a privacy protection scheme $\mathcal{M}$ to protect $X$ (see Fig. \ref{protectability-demo}(b)). To evaluate to what extent $\mathcal{M}$ works, we further define Level-of-Protection (Section~\ref{sec_LoP}) on a privacy protection scheme $\mathcal{M}$. Intuitively, $\mathcal{M}$ will perturb the feature set $Z$. This operation changes the contributions of the features and protects the features. Accordingly, we will define protected features $Z_D$. Our Level-of-Protection metric, the $\mathcal{LP}$-score, is defined as the ratio of protected features against the full feature set.


\subsection{Protectable Features}\label{sec_protectable}
Protectable features are determined according to the contribution that they make to infer private attribute $Y_{pri}$ and target analytics attribute $Y_a$. To achieve this, we need to first quantify the contribution of each feature to both $Y_{pri}$ and $Y_a$. 

For the contribution of each feature $z_i\in Z$, we define it with the predictive power function $\nu_F$ by quantifying how critical $z_i$ is to the accurate prediction of model $F$.\footnote{There are measures other than predictive power, e.g., Grad-CAM~\cite{Selvaraju2017}, LIME~\cite{Ribeiro2016}. We plan to investigate alternatives in our future work.} There are several methods proposed to quantify the feature contribution with the predictive power function, such as feature ablation~\cite{Jing2018} and Bivariate association~\cite{Yao-Zhong2009} which are measured by assuming each feature is independent. However, feature correlation exists in the prediction, and we adopt Shapely value~\cite{Shapley1952} to assign the contribution score to each feature fairly. 
Specifically, each feature $z_i$ participates in a coalition $S\subseteq Z$ to collaboratively make contributions to the decision of the deep learning model $F$, and the predictive power function $\nu_F$ can be seen as the game function. Let $\mu_{F}(z_i, S, Y_a)$ be the payoff for feature $z_i$ in the coalition $S$, then we can define it as the predictive power difference when $z_i$ participate in $S$, and we have  $\mu_{F}(z_i, S, Y_a)=\nu_F(S\cup {z_i}, Y_a)-\nu_F(S, Y_a)$. As there are many potential combinations of feature set $S$, the contribution score of feature $z_i$ can be defined as the expected mean of predictive power difference over all the combinations of $S$. 

\begin{definition}\label{contribution_score}
(Feature contribution score) Given data matrix $X$ and $Z$ is the feature set extracted from $X$ for predicting the attribute $Y_a$ with model $F$. The contribution score of each feature $z_i\in Z$ is:
\begin{equation}\label{feature_con}
    \begin{aligned}
       \mathcal{C}(z_i, Y_a)= \sum_{S \subseteq Z \backslash\{z_i\}} w_S\cdot \mu_{F}(z_i, S, Y_a),
    \end{aligned}
\end{equation}
where
\begin{equation}
    \begin{aligned}\label{weight_s}
        w_S=\frac{|S|!(|Z|-|S|-1)!}{|Z|!},
    \end{aligned}
\end{equation}
and $S$ is the set of features excluding feature $z_i$.
\end{definition}

With Definition~\ref{contribution_score}, we can define the contribution scores of feature $z_i$ to attribute $Y_{pri}$ as $\mathcal{C}(z_i, Y_{pri})$. Then, we can determine whether the features are protectable based on their contribution scores. Specifically, the feature $z_i$ is protectable if it contributes none or little to predicting $Y_{pri}$. Let $\epsilon$ be the privacy budget, which represents the required minimum contribution that an attacker can infer the private attribute $Y_{pri}$. Then the protectable feature can be defined as the feature of which the contribution to $Y_{pri}$ is less than $\epsilon$, and we have the following definition.
\begin{definition}\label{protectable_de}
(Protectable features) Let $Z$ be the feature set extracted from data $X$ for predicting the target attribute $Y_a$, and $Y_{pri}$ be the private attribute of data $X$. The protectable features $Z_P$ is given by:
\begin{equation}\label{protectable_eq}
    \begin{aligned}
        Z_P=\{z_i|\mathcal{C}(z_i, Y_{pri}) \leq \epsilon, z_i \in Z\},
    \end{aligned}
\end{equation}
where $\epsilon$ is a constant threshold.
\end{definition}

\subsection{Privacy Protectability and The Estimation Algorithm}\label{sec_protectability}
In this subsection, we first introduce the definition of the proposed privacy protectability based on the definition of protectable features, and then design an algorithm to compute the privacy protectability efficiently.

\textbf{Privacy Protectability Definition.} The underlying philosophy of perturbation-based protection schemes is to reduce the sensitive information contained in the exposed feature set while guaranteeing the accuracy of the analytics task. This trade-off may work when most of the features are protectable and may fail when protectable features make a few contributions for predicting the target attribute $Y_a$. To evaluate whether perturbation-based protection schemes work, a specific scheme should be implemented and applied first, and then its protection performance is judged by calculating the accuracy of the inference attack and analytics task with a test video. However, this process includes intensive computations which may lead to significant time costs in practice. Therefore, to efficiently measure when will a privacy protection scheme work, we propose a new metric called privacy protectability, the $\mathcal{P}$-score, by analyzing to which degree the contributions made by protectable features are preserved.

Intuitively, the more contributions for predicting the target attribute $Y_a$ made by the protectable features, the higher probability that a privacy protection scheme may work. To reflect this, we can define privacy protectability based on the proportion of contributions made by protectable features with respect to the whole set of features.
\begin{definition}\label{proctectionability_de}
(Privacy protectability) Let $Z_P \subseteq Z$ be the set of protectable features and $Z$ be the full feature set of $X$ extracted for predicting attribute $Y_a$. The privacy protectability of $Z$ with respect to the private attribute $Y_{pri}$ of data $X$, denoted by $\mathcal{P}(Z,Y_a, Y_{pri})$, is measured by the proportion between the contributions of $Z_P$ and $Z$: 
\begin{equation}\label{proctectionability_eq}
    \begin{aligned}
    \mathcal{P}(Z, Y_a, Y_{pri})=\frac{\sum_{z_i\in Z_P}\mathcal{C}(z_i, Y_a)}{\sum_{z_j\in Z}\mathcal{C}(z_j, Y_a)},
    \end{aligned}
\end{equation}
where $Z_P$ satisfies Eq.~(\ref{protectable_eq}).
\end{definition}

With privacy protectability defined, we consider how to compute it efficiently. The main computation overhead comes from calculating the feature contribution score which has two challenges. To overcome these, we design an $\mathcal{P}$-score estimation algorithm in the following.

\textbf{The Privacy Protectability Estimation (PPE) algorithm.} The key to computing $\mathcal{P}$-score is the estimation of the feature contribution score. There are two challenges to estimating the feature contribution score. First, the predictive power $\nu_F(S, Y_a)$ needs to calculate the loss. This requires executing the DNN model $F$ so as to compare the ground truth with the DNN model outputs; this is computation-intensive. Second, the computing of the weight $w_S$ introduced by the Shapely value has exponential complexity. 

To address the first challenge, we note that we can approximate the loss by executing the DNN model partially; as demonstrated in \cite{compression}, the loss can be computed without executing the complete model. We leave it as a future work to study the most appropriate approximation; and in this paper, we adopt an approximation focusing on algorithmic efficiency. This naturally reduces to using mutual information to approximate the predictive power function. This also reflects that most loss function of analytics tasks is the cross entropy or mean squared error (MSE). Specifically, suppose $F$ is the optimal prediction model for analyzing $Y_a$, and we have $\nu_{F}(S, Y_a)= I(Y_a; S)$ when the loss function is cross entropy or MSE~\cite{Covert2020}. Here, $I$ denotes the mutual information. The feature contribution score in Eq.~(\ref{feature_con}) can be rewritten as $\mathcal{C}(z_i, Y_a)= \sum_{S \subseteq Z \backslash\{z_i\}}w_S\cdot I(Y_a; z_i|S)$,
where $I(Y_a;z_i|S)=I(Y_a;S,z_i)-I(Y_a; S)$.

To address the second challenge, i.e., the weight $w_S$ based on Shapely value has computation exponentially increased with the number of features increased, we adopt the Monte Carlo sampling method to randomly sample several feature set instead of all combinations as in~\cite{Covert2020}. Specifically, let $\{S_1, S_2, \ldots, S_M\}$ be the feature subsets randomly sampled from the uniform distribution of full set $Z$. Then, for each feature subset $S_m$, we can compute its weight $w_{S_m}$ according to Eq.(\ref{weight_s}). With the sampled weight $w_{S_m}$ and the approximated predictive power obtained, we can derive the contribution score of each feature $z_i$.

There are three steps in estimating the $\mathcal{P}$-score in Algorithm~\ref{algorithm_1}. We first estimate the contribution score of each feature for inferring the target analytics attribute $Y_a$ and the private attribute $Y_{pri}$ (Line 5-8). Then, we construct the protectable features $Z_P$ from the full feature set $Z$, by selecting features whose contribution scores for inferring the private attribute $Y_{pri}$ are less than a threshold $\epsilon$ (Line 11). Finally, we obtain the $\mathcal{P}$-score by calculating the proportion of total contribution scores of the protectable feature set $Z_P$ with respect to the full feature set $Z$ (Line 16). 

\begin{algorithm}[tb]
    \caption{Privacy Protectability Estimation (PPE)}
    \label{algorithm_1}
    \textbf{Input}: $X$, $Z$, $F$, $\epsilon$, $M$, $Y_a$, $Y_{pri}$\\
    \textbf{Output}: $\mathcal{P}(Z, X, Y_a, Y_{pri})$
    \begin{algorithmic}[1] 
        \STATE Let $Z_P=\emptyset$;
        \FOR{$i=1,2,\ldots,n$}
            \STATE $\mathcal{C}(z_i, Y_a)\leftarrow 0$, $\mathcal{C}(z_i, Y_{pri})\leftarrow 0$, $\mathcal{M}\leftarrow\{\emptyset,\ldots,\emptyset\}$;
            \STATE \text{// Step I: Feature contribution score estimation}
            \FOR{$m=1,2,\ldots,M$}
            \STATE $S_m \leftarrow$Sample from $Z\backslash\{z_i\}$;
            \STATE $w_{S_m}= \frac{|S_m|!(|Z|-|S_m|-1)!}{|Z|!}$, $\mu_F(z_i, S_m, Y_a)= I(Y_a; z_i|S_m)$, $\mu_F(z_i, S_m, Y_{pri})= I(Y_{pri}; z_i|S_m)$;
            \STATE $\mathcal{C}(z_i, Y_a)= \sum_{m=1}^M w_{S_m}\cdot\mu_F(z_i, S_m, Y_a)$, $\mathcal{C}(z_i, Y_{pri})= \sum_{m=1}^M w_{S_m}\cdot\mu_F(z_i, S_m, Y_{pri})$;
            \STATE \text{// Step II: Protectable feature selection}
            \IF{$\mathcal{C}(z_i, Y_{pri})\leq \epsilon$}
                \STATE $Z_P\leftarrow z_i$;
            \ENDIF
            \ENDFOR
        \ENDFOR
        \STATE \text{// Step III: Privacy protectability calculation}
        \STATE $\mathcal{P}(Z, Y_a, Y_{pri})=\frac{\sum_{z_i\in Z_P}\mathcal{C}(z_i, X, Y_a)}{\sum_{z_j\in Z}\mathcal{C}(z_j, Y_a)}$;
        \STATE \textbf{return} $\mathcal{P}(Z, Y_a, Y_{pri})$.
    \end{algorithmic}
\end{algorithm}

\subsection{Level-of-Protection}\label{sec_LoP}
For a given analytics task, a video stream that has high privacy protectability indicates that it has a high probability of being protected. That is, there exists a perturbation-based privacy protection scheme to make the video stream protected. We define level-of-protection, $\mathcal{LP}$-score, to quantify to what extent a protection scheme works.

Let $\mathcal{M}$ be a privacy protection scheme. For each feature $z_i\in Z$, it is transformed to $\hat{z}_i=\mathcal{M}(z_i)$ when the protection scheme is applied. Let $\mathcal{C}_p(\mathcal{M}, z_i, Y_{pri})$ be the preserved contribution score of $z_i$ to predicting $Y_{pri}$ after perturbing by $\mathcal{M}$, we have
\begin{equation}
    \begin{aligned}
        \mathcal{C}_p(\mathcal{M}, z_i, Y_{pri}) = \mathcal{C}(\hat{z}_i, Y_{pri}).
    \end{aligned}
\end{equation}
Then, the preserved contribution of $z_i$ to $Y_a$ after perturbing by $\mathcal{M}$ is $\mathcal{C}_p(\mathcal{M}, z_i, Y_a)$. We define a feature $z_i$ as a \textit{protected feature} when its preserved contribution $\mathcal{C}_p(\mathcal{M}, z_i, Y_{pri})$ to private attribute is less than the threshold $\epsilon$. 

\begin{definition}\label{protected_de}
(Protected features) Let $\mathcal{M}$ be a privacy protection scheme, $Z$ be the feature set extracted from data $X$ for predicting the target attribute $Y_a$, and $Y_{pri}$ be the private attribute of data $X$. With the protection scheme $\mathcal{M}$ is applied, the protected features $Z_D$ is given by:
\begin{equation}\label{protected_eq}
    \begin{aligned}
        Z_D=\{z_i|\mathcal{C}_p(\mathcal{M}, z_i, Y_{pri}) \leq \epsilon, z_i \in Z\}
    \end{aligned}
\end{equation}
where $\epsilon$ is a constant threshold.
\end{definition}

Then, we define the $\mathcal{LP}$-score as the proportion of preserved contributions made by protected features $Z_D$ with respect to the whole set of perturbed features $Z$.

\begin{definition}\label{level-of-protection_de}
(Level-of-protection) Let $\mathcal{M}$ be a privacy protection scheme and $Z_D\subseteq Z$ be the set of protected features. The level-of-protection of $\mathcal{M}$ on the private attribute $Y_{pri}$ of data $X$, denoted by $\mathcal{LP}(\mathcal{M}, Z,Y_a, Y_{pri})$, is measured by the proportion between the preserved contributions of $Z_D$ and $Z$: 
\begin{equation}\label{proctectionability_eq}
    \begin{aligned}
    \mathcal{LP}(\mathcal{M}, Z, Y_a, Y_{pri})=\frac{\sum_{z_i\in Z_D}\mathcal{C}_p(\mathcal{M}, z_i, Y_a)}{\sum_{z_j\in Z}\mathcal{C}_p(\mathcal{M}, z_j, Y_a)},
    \end{aligned}
\end{equation}
where $Z_D$ satisfies Eq.~(\ref{protected_eq}).
\end{definition}
The computation of $\mathcal{LP}$-score is similar to privacy protectability, and we can follow Algorithm~\ref{algorithm_1} to estimate $\mathcal{LP}$-score by replacing the feature value $z_i$ with the feature value $\hat{z}_i$ perturbed by $\mathcal{M}$.

\section{Evaluation}\label{eval}

Our evaluation tries to answer two questions: (1) How does the privacy protectability metric reflect the empirical performance of privacy protection schemes? We validate this by using a set of benchmarking privacy protection schemes and examining how our privacy protectability metric correlates with their empirical performance. (2) How does our Privacy Protectability Estimation (PPE) algorithm perform? We examine how PPE approximates the feature contribution score defined by predictive power and its computational efficiency.

\subsection{Evaluation Setup}
\textbf{Datasets.} We measure the performance of privacy protectability on CelebA~\cite{liu2015faceattributes}, which is a face attributes dataset of over 200K face images with 40 attribute annotations. And we choose gender as the private attribute.



\textbf{Analytics model.} We validate privacy protectability with six analytics tasks. Generally, each analytics task is supported by a model trained with ResNet-18~\cite{Kaiming2016}.


\textbf{Attack model.} We implement a gender attribute inference attacker, which is a classifier sharing the same model architecture with the task model as in~\cite{Jinyuan2018}.


\textbf{Privacy protection schemes.} We select three representative privacy protection schemes as bench-marking schemes: 1) Differential privacy (DP)~\cite{liu2021dp} adds Gaussian noise to prevent privacy leakage; 2) Shredder~\cite{Mireshghallah2020} learns appropriate noise to add to prevent privacy leakage; 3) DISCO~\cite{Singh2021} obfuscates sensitive information by a data-driven pruning filter.


\textbf{Baselines.} We compare privacy protectability with \textit{Empirical privacy protection performance (EP)}. Let $\mathcal{H}$ be a set of privacy protection schemes. Given a video analytics task, a video stream, and an attribute inference attack, each $H_i\in\mathcal{H}$ will have its task accuracy $Acc_a^i$ and attack accuracy $Acc_{pri}^i$. Intrinsically, $\mathcal{P}$-score reflects the utility-privacy trade-off. For a direct comparison with $\mathcal{P}$-score, we define a performance metric $EP_i$ for $H_i$ as $\frac{Acc_a^i}{Acc_{pri}^i}$. Since the maximal value among all schemes should be closer to the optimal performance reflected by the $\mathcal{P}$-score, we define the empirical performance $EP$ on $\mathcal{H}$ as $max_{H_i\in\mathcal{H}} EP_i$. Moreover, we compare the PPE algorithm with \textit{Model Loss (MLoss)} which follows the same pipeline of the PPE algorithm except that it directly uses the analytics model loss to compute the feature contribution score, i.e., using Eq.~(\ref{feature_con}) in line 7 of PPE algorithm.

\begin{figure}[t]
	\subfigure[Utility-Privacy]{
		\label{two_tasks_tradeoff}
		\includegraphics[scale=0.26]{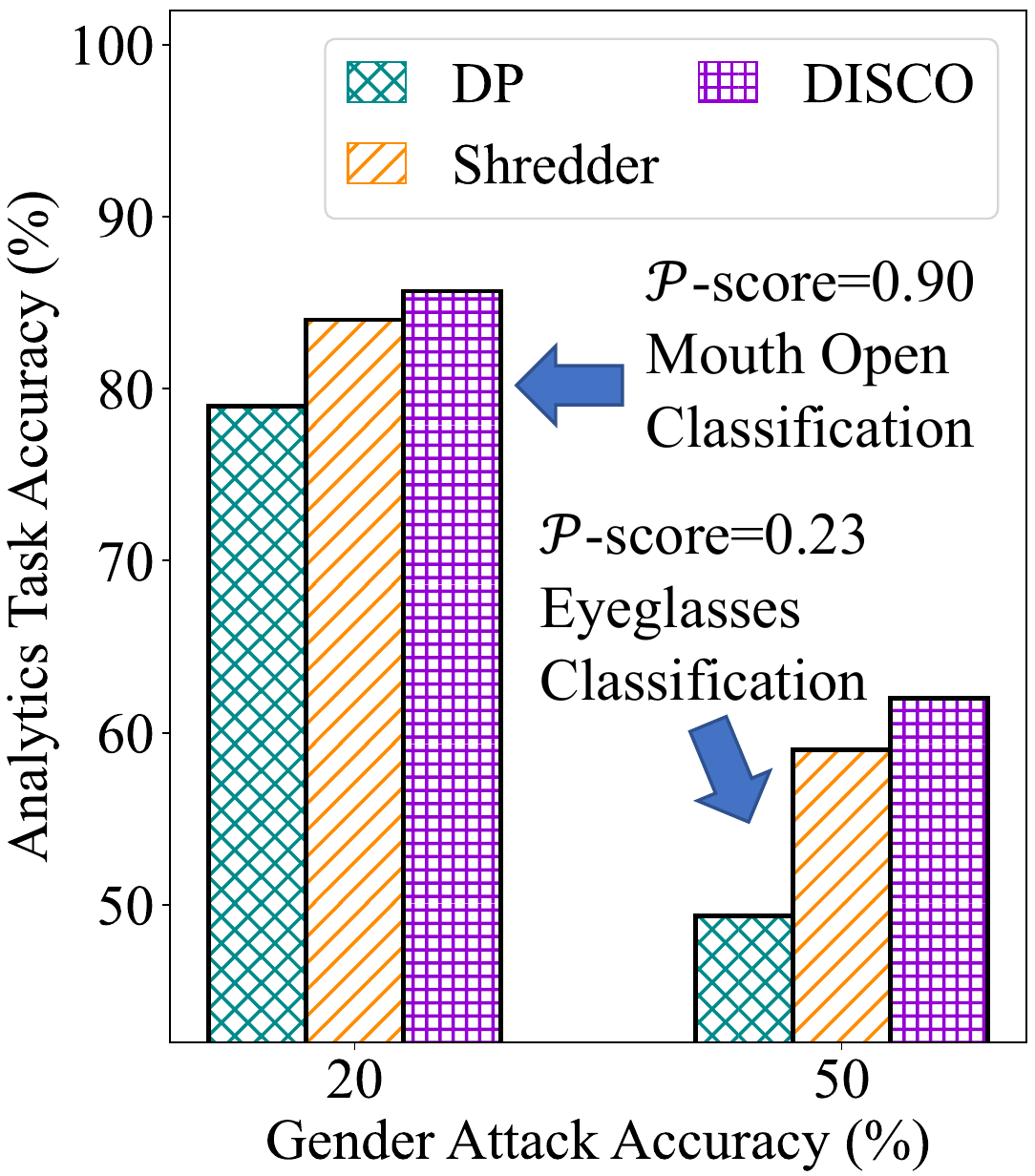}}
 	\subfigure[Visualization]{
		\label{two_tasks_visual}
		\includegraphics[scale=0.24]{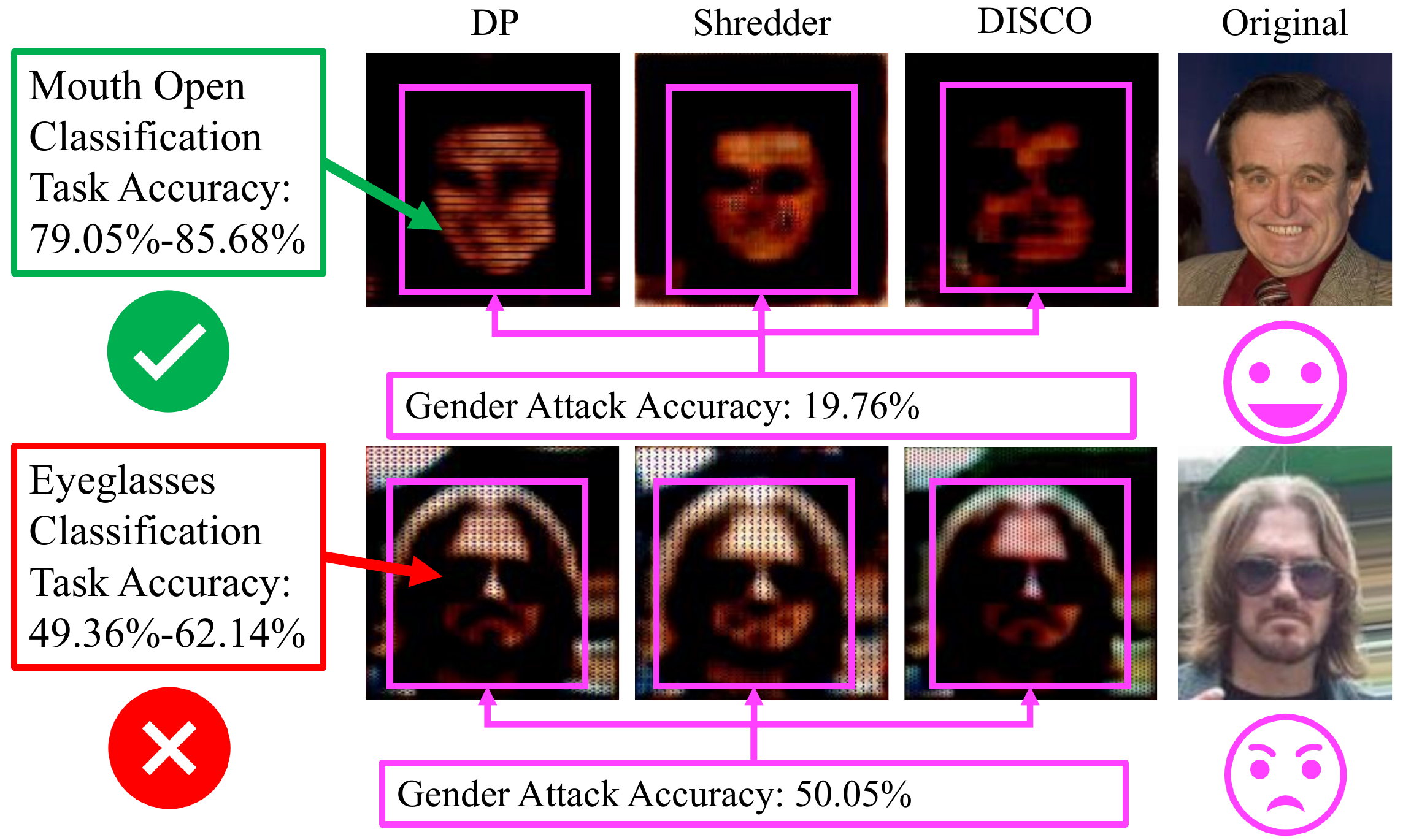}}
	\caption{Comparison of high and low protectability VA tasks.}
	\label{validation_2tasks}
\end{figure}

\subsection{Validation of the Privacy Protectability Performance}


In Fig.~\ref{two_tasks_tradeoff}, we validate the performance of privacy protectability with the eyeglasses classification task and the mouth open classification task on dataset CelebA, and the privacy attribute is gender. We demonstrate the empirical privacy protection performance with DP, Shredder, and DISCO. The mouth open classification task can be well protected with the average attack accuracy being 19.76\% and the average analytics accuracy being 82.37\%, which also can be reflected by the high $\mathcal{P}$-score 0.90. In contrast, the eyeglasses classification task is not protected, with the average attack accuracy being 50.05\% and the average analytics accuracy being 55.75\%. The $\mathcal{P}$-score reflects this with a low value of 0.23. We also visualize these two tasks in Fig.\ref{two_tasks_visual} with the three privacy protection schemes applied. For the upper mouth open classification task, we can classify the mouth as opened with up to 85.68\% accuracy, while the gender of the person is hard to tell. For the lower eyeglasses classification task, the eyeglasses are hard to classify with less than 62.14\% accuracy, but the gender is easy to identify. Therefore, our evaluation shows that privacy protectability can effectively reflect the empirical performance of privacy protection schemes. 


\begin{table*}[!t]
	\centering
	\begin{minipage}[b]{0.645\textwidth}
		\small
		\caption{Correlation between empirical performance of privacy protection schemes and $\mathcal{P}$-score (Pearson correlation coefficient is 0.99)}
        \centering
        \resizebox{\linewidth}{!}{
        \begin{tabular}{ccccccc}
        \hline
        \diagbox[width=10em]{\textbf{Metric}}{\textbf{Analytics Task}} & \textbf{Eyeglasses} & \textbf{Smiling} &\textbf{Wearing Hat} &\textbf{Black Hair} & \textbf{Big Nose}  &\textbf{Mouth Open} \\ \hline
        \textbf{EP} & 1.08 & 1.96 & 2.43 & 2.67 & 3.26  & 4.17 \\
        \textbf{$\mathcal{P}$-score} & 0.23 & 0.42 & 0.49 & 0.53 & 0.69 & 0.90\\
        
        \hline
        \end{tabular}
        }
    \label{validation-correlation}
	\end{minipage}\quad
	\begin{minipage}[b]{0.325\textwidth}
		\flushright
		\small
		\caption{Effectiveness comparison of PPE with MLoss for $\mathcal{P}$-score.}
		\resizebox{\linewidth}{!}{
		\begin{tabular}{ccccc}
        \hline
        \diagbox[width=10em]{\textbf{Metric}}{\textbf{Sample Size}} & \textbf{50} & \textbf{100} & \textbf{150} & \textbf{200} \\ \hline
        \textbf{PPE} & 0.90 & 0.92 & 0.89 & 0.91 \\
        \textbf{MLoss} & 0.92 & 0.90 & 0.88 & 0.91 \\
        
        \hline
        \end{tabular}
		}
    \label{validation-table}
	\end{minipage}\quad
\end{table*}




We measure the $\mathcal{P}$-scores and EP of six classification tasks under three privacy schemes to quantify the correlation between empirical privacy protection performance and privacy protectability. Results are shown in Table~\ref{validation-correlation}. We can see that a higher $\mathcal{P}$-score corresponds to a higher EP. For example, the $\mathcal{P}$-score of the mouth open classification task is the highest at 0.90, and its EP is 4.17, which is also the highest. We compute the Pearson correlation coefficient to quantitatively measure the correlation between EP and $\mathcal{P}$-score, and the value is 0.99, indicating a strong correlation between them. According to this evaluation, the $\mathcal{P}$-score can be directly utilized to represent empirical privacy protection performance.

\subsection{Performance of the Privacy Protectability Estimation Algorithm}

\textbf{Effectiveness analysis.} We further evaluate the effectiveness of the PPE algorithm by comparing it to MLoss on the mouth open task with four different sample sizes $M=\{50, 100, 150, 200\}$. Table~\ref{validation-table} shows the results. MLoss represents the optimal measure of privacy protectability as it is directly computed based on the predictive power function. We can see the $\mathcal{P}$-score of PPE is almost the same as the MLoss, verifying that our approximation is effective. Moreover, the $\mathcal{P}$-score is around 0.90 for all sample sizes, indicating that sample size will not significantly affect our measure .






\textbf{Efficiency analysis.} We measure the efficiency of PPE by comparing the computation cost to MLoss in terms of floating point operations (FLOPs) with four sample sizes $M=\{50, 100, 150, 200\}$. According to Table~\ref{efficiency-table}, we can see that the computation cost of both PPE and MLoss grows linearly as the sample size increases. Nevertheless, for the same sample size, the computation overhead of MLoss is up to 1062 times larger than PPE. The underlying reason for this significant gap is execution of the computation-intensive DNN model. Therefore, the evaluation results validate that our approximation PPE algorithm is more efficient in practice.


\begin{table}[t]
\caption{Efficiency comparison of PPE with MLoss (Unit: TFLOPs) }
\label{efficiency-table}
\centering
\normalsize
\resizebox{\linewidth}{!}{
\begin{tabular}{ccccc}
\hline
\diagbox[width=10em]{\textbf{Metric}}{\textbf{Sample Size}} & \textbf{50} & \textbf{100} & \textbf{150} & \textbf{200} \\ \hline
\textbf{PPE} & 0.85  & 1.70  & 2.56 & 3.42  \\
\textbf{MLoss} & 902.40 & 1804.80 & 2707.21 & 3609.61 \\
\hline
\end{tabular}
}
\end{table}

\section{Case Study}

We now present a case study on using privacy protectability metrics in practice. We study how to conduct privacy protection for two example tasks, a traffic light detection task and a crowd density detection task in open areas. We study conventional approaches and an approach assisted by the privacy protectability metrics developed in this paper. 

Currently, examining whether an analytics task can apply a certain privacy protection scheme to a video stream requires a few steps. Specifically, after choosing a privacy leakage budget $\epsilon$, one can (1) apply the protection scheme on the video stream to obfuscate sensitive information; (2) conduct the video analytics task on the obfuscated video stream; and (3) if the analytics accuracy is less than what is acceptable, one may choose to evaluate another protection scheme, or one may choose to apply encryption-based schemes, etc.\footnote{In practice, there may be no definitive privacy leakage budget; and one can calibrate the privacy leakage thresholds and the analytics accuracy by iterating the aforementioned steps.}
Such a procedure brings about significant computation overheads.

In comparison, we can design a new procedure with our privacy protectability metrics. Specifically, one first needs to choose a privacy protectability threshold $Th_{P}$, representing how willing one is to trade reduction of privacy leakage with the sacrifice of utility. For example, if one cares about task accuracy and privacy leakage equally, this threshold should be large, and he can set it empirically, e.g., as 0.7. And the violation of $Th_{P}$ indicates that an encryption-based scheme should be applied. In this paper, we leave a full-ranged investigation on setting this threshold for future work.

After choosing a privacy leakage budget $\epsilon$ and a privacy protectability threshold $Th_{P}$, one can (1) apply the PPE algorithm on the video stream; (2) if $\mathcal{P}$-score $<Th_{P}$, apply an encryption-based scheme (or choose another video stream to complete the video analytics task); otherwise, (3) apply the privacy protection scheme with privacy leakage budget $\epsilon$ to obfuscate the video stream and compute $\mathcal{LP}$-score, which reflects the accuracy of the video analytics task. For a well-designed privacy protection scheme, we may expect the $\mathcal{LP}$-score to be acceptable.\footnote{If $\mathcal{LP}$-score is not acceptable, we can choose another privacy protection scheme.} Note that in this new approach, the computation of the $\mathcal{P}$-score and $\mathcal{LP}$-score are all intrinsically sampling-based, which is significantly less resource-consuming than executing the video analytics model.

We choose a real-world video captured along 6th Avenue in New York.\footnote{\url{https://www.youtube.com/watch?v=kBleVYw_MDs}} Based on where the car passed, the video is cut into three clips: West-12, West-26 and, West-34, starting at West 12th, West 26th, and West 34th street, respectively. We implement the traffic light detector and crowd density detector with YOLOv3~\cite{mmdetection} on the LISA dataset \cite{lisa} and the WiderPerson dataset \cite{widerperson}, respectively. A face detector trained by YOLOv3 on the WiderFace dataset~\cite{widerface} is used as the attack model to infer if any face exists in a frame.


\begin{table*}[!t]
	\centering
	\begin{minipage}[b]{0.32\textwidth}
		\small
		
		\caption{Comparison of $\mathcal{P}$-score.}
        \resizebox{\linewidth}{!}{
        \begin{tabular}{cccc}
        \hline
        \diagbox[width=10em]{\textbf{Analytics task}}{\textbf{Video}}
          &West-12&West-26&West-34\\
        \hline
        \textbf{Traffic Light} & 0.75 & 0.76 & 0.74 \\
        \hline
        \textbf{Crowd Density} & 0.29 & 0.31 & 0.27 \\
        \hline
        \end{tabular}
        }
    \label{case-table1}
	\end{minipage}\quad
	\begin{minipage}[b]{0.63\textwidth}
		\flushright
		\small
		
		\caption{Comparison of $\mathcal{LP}$-score. }
		\resizebox{\linewidth}{!}{
        \begin{tabular}{cccccccccc}
        \hline
        \multirow{2}*{\diagbox[width=10em]{\textbf{Analytics task}}{\textbf{Video}}}
          &\multicolumn{3}{c}{West-12} & \multicolumn{3}{c}{West-26}& \multicolumn{3}{c}{West-34}\\
          \cline{2-10}
          & DP & Shredder & DISCO & DP & Shredder & DISCO & DP & Shredder & DISCO\\
        \hline
        \textbf{Traffic Light} & 0.78 & 0.78 & 0.87 & 0.77 & 0.79 & 0.88 & 0.77 & 0.79 & 0.88\\
        \hline
        \textbf{Crowd Density} & 0.46 & 0.47 & 0.53 & 0.45 & 0.47 & 0.52 & 0.44 & 0.48 & 0.54\\
        \hline
        \end{tabular}
        }
    \label{case-table2}
	\end{minipage}\quad
\end{table*}

\begin{table}[t]
\caption{Computation cost of conventional and protectability-assisted privacy protection scheme evaluation (Unit: TFLOPs) }
\label{case-cost}
\centering
\resizebox{\linewidth}{!}{
\begin{tabular}{ccccc}
\hline
\diagbox[width=10em]{\textbf{Procedure}}{\textbf{Scheme}} & \textbf{DP} & \textbf{Shredder} & \textbf{DISCO} & \textbf{Total} \\ \hline
\textbf{Conventional} & 1185.32 & 1185.32 & 1488.61 & 3859.25  \\
\textbf{Protectability-assisted} & 0.70 & 0.70 & 34.30 & 35.70  \\
\hline
\end{tabular}
}
\end{table}

The $\mathcal{P}$-score of three video streams for two tasks is shown in Table~\ref{case-table1}. The traffic light detection task with three videos has a higher $\mathcal{P}$-score at around 0.75, while the $\mathcal{P}$-score of the crowd density detection task is below 0.31. It indicates that the crowd density detection task is unprotectable with the perturbation-based privacy protection scheme, which is caused by the high correlation between features required by the crowd density detection and the human face attack.
Next, we evaluate the privacy protection performance by applying DP, Shredder, and DISCO on the two tasks with three video clips, and the corresponding $\mathcal{LP}$-scores are shown in Table~\ref{case-table2}. We can see the unprotectable crowd density detection task indeed has a low $\mathcal{LP}$-score with all privacy protection schemes, which is around 0.44 to 0.54. In contrast, the $\mathcal{LP}$ for the protectable traffic light detection task is much higher with an average value of 0.80. Particularly, DISCO outperforms the other two schemes for all three videos, and thus it can be chosen to apply to protect the video.

Table~\ref{case-cost} shows the efficiency of two procedures by comparing the computation cost of applying DP, Shredder, and DISCO separately and the total cost. For the conventional procedure, the computation cost consists of applying the privacy protection scheme and executing the video analytics task. For the protectability-assisted procedure, the computation cost includes calculating $\mathcal{P}$-score, applying the privacy protection scheme, and computing $\mathcal{LP}$-score. The computation cost of the traditional approach is about 1143 times higher than our protectability-assisted procedure on average. The underlying reason for this gap is that the sample-based PPE algorithm does not require running the computation-intensive analytics DNN model. The cost of our approach for DISCO is higher than for the other two schemes because DISCO needs DNN-based pruning at run-time, while DP and Shredder only perform simple noise addition.

\section{Related Work}

\textbf{Inference Privacy.} Privacy-preserving machine learning (PPML) can be classified into training privacy and inference privacy~\cite{xu2021privacy,mireshghallah2020privacy,Jegorova2022,Al-Rubaie2019}. Training privacy focuses on protecting the sensitive information in the training dataset. There are membership inference attacks \cite{Jiayuan2022}, property inference attacks \cite{carlini2019secret}, model inversion attacks \cite{Ziqi2019}, etc; and protection schemes include differential privacy \cite{Martin2016}, adversarial training~\cite{Jinyuan2019}, etc.

Our study falls into inference privacy, where the privacy of the inference data needs to be protected. Attribute inference attack \cite{Shagufta2022} is one important inference attack that reveals sensitive attributes. Another inference attack, the reconstruction attack~\cite{dosovitskiy2016inverting}, tries to recover the input image with the hijacked intermediate results of the model. One privacy protection approach attempts to construct a secure environment, i.e., to ensure that the data can only be accessed or decoded by trusted parties, e.g., homomorphic encryption~\cite{Reagen2021}, trusted execution environment \cite{narra2019privacy}, etc, which is resource-demanding. Another privacy protection approach tries to perturb the inference data to reduce sensitive information leakage while preserving analytics task accuracy. For example, differential privacy~\cite{liu2021dp} obfuscates data, and DISCO~\cite{Singh2021} prunes the sensitive channels.

Our work is orthogonal to all privacy protection schemes. Privacy protectability characterizes the intrinsic property of whether data can be perturbed in a way that preserves both privacy and analytics accuracy. From the viewpoint of privacy protectability, the schemes construct a secure environment make data protectable by data access control.






\textbf{Privacy-preserving Video Analytics Systems.}
Privacy protection attracts increasing investigations in video analytics systems~ \cite{priconcern}. One particular factor for video analytics systems is their real-time delay requirements. With resource constraints in the edge, existing privacy-preserving video analytics systems emphasize light-weight perturbation-based schemes, e.g., injecting noises (i.e., NoPeek~\cite{nopeek2020}), transforming or removing the sensitive information (i.e., Preva~\cite{Rui2022}), in the transmitted data. However, the performance of a privacy protection scheme is usually measured by empirical experiments, which are resource-consuming and it is difficult to have a definitive answer on whether there exists a privacy protection scheme to perform well. As a matter of fact, there are reports that privacy protection schemes do not work in certain scenarios \cite{Singh2021}.

Our privacy protectability provides a quantitative measure on whether a perturbation-based privacy protection scheme can protect a video data. We also define a level-of-protection metric to measure to what extent a privacy protection scheme works. Our metrics can be efficiently computed, allowing operational decisions to be made in practice.

\section{Conclusion}

In this paper, we proposed a new metric privacy protectability for inference privacy. Privacy protectability characterizes to what degree a video stream can be protected  given a certain video analytics task. It is defined through information theory, where we quantified the contribution of the features to the video analytics accuracy and to the privacy leakage; and to what degree they can be separated. Privacy protectability lays the foundations for understanding whether a privacy protection scheme can work and to what extent it works. It also naturally assists decision-making in practice on how to apply protection schemes, as we demonstrated in a case study. 





\bibliographystyle{named}
\bibliography{ijcai23}

\end{document}